\documentstyle[graphicx]{article}

\begin{document}

\title{Results on Deeply Virtual Compton Scattering \\
at Jefferson Lab}

\author{F.~Sabati\'e}

\maketitle

\begin{abstract}
After about 10 years of growing interest for Generalized Parton Distributions come the first results from dedicated experiments, using
the golden Deeply Virtual Compton Scattering process. After a short introduction, we will explain the experimental methodology and 
show results of the Hall A E00-110 experiment, which aimed at measuring helicity-dependent photon electroproduction cross sections.
We will enphasize how this experiment provided the first stringent tests of the scaling property of this process, allowing for the
first time a model-independent extraction of a linear combination of Generalized Parton Distributions.
We will also describe the Hall B E01-113 experiment which
measured the photon electroproduction beam spin asymmetry over a wide kinematical range. The summary will include an outlook on
the next generation of experiments which are already planned at Jefferson Lab at 6~GeV, but also after the planned 12~GeV upgrade.
\end{abstract}


\section{Introduction}

These are very exciting times for the field of Generalized Parton Distributions (GPD). Since their theoretical
introduction in the mid 90's \cite{Mu,Rad1,Rad2,Ji1,Ji2,Diehl1},
a handful of non-dedicated results came from HERA and Jefferson Lab \cite{Adloff,Chekanov,Step,Aira,Chen},
showing that Deeply Virtual Compton Scattering is potentially a very important tool for the understanding of the nucleon structure for the years
to come. A lot of theoretical progress has been made over the last 10 years: the full harmonic structure of the electroproduction cross section has
been calculated up to twist-3 \cite{Bel1}, interpretation of GPDs in the transverse plane, either at 0 or finite skewdness
\cite{Burk,Diehl2}, nuclear GPDs \cite{Freund,Cano}, and many other topics \cite{Diehl3,Bel2}.

After being proposed in the years 2000-2003, three dedicated experiments ran at Jefferson Lab in 2004-2005:
\begin{itemize}
\item E00-110 in Hall A measured helicity-dependent photon electroproduction cross sections, aiming at checking the factorization theorem in the Jefferson Lab
energy range, and making the first measurement of GPDs. We will give details about this experiment in the next section.
\item E03-106 in Hall A is the almost the same experiment as E00-110 but on the neutron, using a deuterium target. Analysis of this experiment is still in progress,
and we will only mention it \cite{Malek}.
\item E01-113 in Hall B was aimed at measuring the beam single spin asymmetry in a wide kinematical range, in order to constrain GPD models as much as possible.
We will give details about this experiment in the third section.
\end{itemize}

The talk given at the symposium as well as these proceedings focus on these dedicated experiments. The following sections will detail the proton experiments E00-110
in Hall A and E01-113 in Hall B of Jefferson Lab. We will briefly mention the future of this GPD program at Jefferson Lab in the conclusion.

\section{Hall A experiment E00-110}

E00-110 was initiated to obtain accurate cross section measurements at different $Q^2$ from 1.5 to 2.3~GeV$^2$ and fixed $x_B=0.36$.
It was designed to ensure a good control on exclusivity, in order to test the hypothesis of twist-2 dominance
in the beam energy range around 6~GeV. The data was acquired in JLab Hall A \cite{Carlos}, using the experimental apparatus described in Fig.~1. The 5.75 GeV
electron beam was incident on a 15~cm liquid H$_2$ target, yielding luminosities of about 10$^{37}$cm$^{-2}$s$^{-1}$ with 76\% beam polarization. The scattered
electron was detected and identified in the left high resolution spectrometer, a standard equipment in Hall A \cite{NimA}. Photons were detected in a lead fluoride electromagnetic
calorimeter, in direct view of the target, located around 110~cm from its center, at angles as low as 15$^\circ$. The signals generated by the PMTs coupled to these
calorimeter blocks were digitized
over 128~ns using VME boards based on the ARS chip \cite{Feinstein}. A trigger was formed between  a good electron in the spectrometer and a high energy cluster above
1~GeV in the calorimeter in order to define an (e,$\gamma$) event. For contamination studies, we used an additional proton array of 100 plastic scintillator blocks
which subtented polar angles from 18 to 38$^\circ$ and azimuthal angles from 45 to 315$^\circ$, in order to detect the proton from the $e p \to e p \gamma$ process
in triple coincidence along with the electron and photon.
\begin{figure}
   \begin{center}  \includegraphics[width=0.8\textwidth]{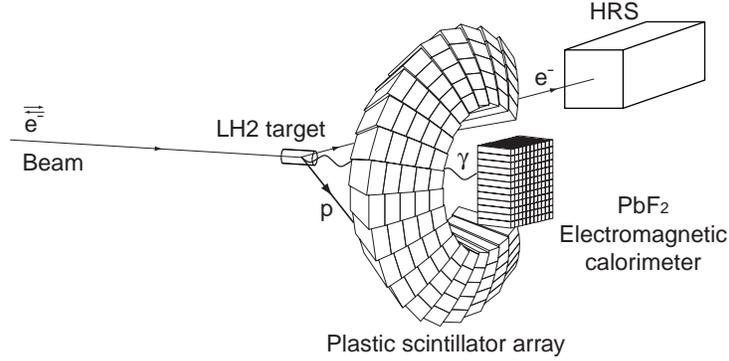}\end{center}
  \caption{Experimental configuration for the E00-110 experiment in Hall A. Scattered electrons were detected in the left High Resolution Spectrometer (HRS),
           photons were detected in a 132-block lead fluoride calorimeter, and a sample of protons were detected in a 100-block plastic scintillator array.}
\end{figure}

The typical H(e,e$\gamma$)X missing mass spectrum is shown in Fig.~2. In order to obtain a pure DVCS
sample, the analysis followed the following procedure: firstly, accidentals were subtracted using a special sample of events. Then using symmetric $\pi^0$ decay in our
calorimeter, we obtained a high statistics H(e,e$\pi^0$)X sample, which was used to infer the asymmetric-decayed $\pi^0$ contamination under our exclusive DVCS
peak in Fig.~2. Finally, the proton array was used to evaluate the contamination from inelastic channels such as $ep\to eN\gamma\pi$: our exclusive sample was
estimated to have less than 3\% contamination from such processes.
\begin{figure}
   \begin{center}  \includegraphics[height=.22\textheight]{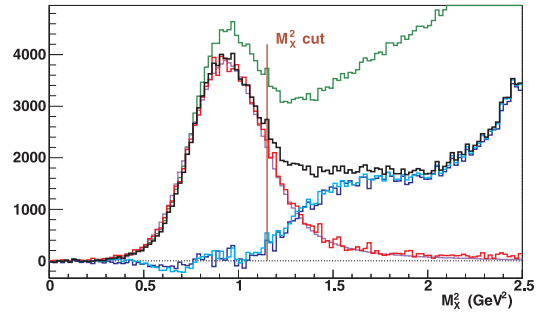}\end{center}
  \caption{Missing mass squared spectrum for H(e,e$\gamma$)X. In green, events with only $(e,\gamma)$ detected. In black, the same events after $\pi^0$
           subtraction as described in the text. In red, $(e,\gamma, p)$ sample using the scintillator array for proton detection, along with the
           corresponding Monte-Carlo prediction in purple. The two very similar blue histograms are generated subtracting the red (or purple) to the black
           histograms, showing the residual contamination of our exclusive sample, estimated to be under 3\%.}
\end{figure}

The difference and total cross sections as a function of the angle between the hadronic and leptonic planes $\phi_{\gamma\gamma}$ can be written
- in the twist-3 approximation - in the following way:
\begin{eqnarray}
d^5\sigma^+ - d^5\sigma^- &=& \Gamma^\Im_1 \Im m \cdot [\mathcal C ^I(\mathcal F)] \cdot \sin\phi_{\gamma\gamma} + \Gamma^\Im_2 \Im m \cdot [\mathcal C ^I(\mathcal F^{eff})]
\cdot \sin 2\phi_{\gamma\gamma} \\
d^5\sigma^+ + d^5\sigma^- &=& \Gamma^\Im_0 \Re e \cdot [\mathcal C ^I(\mathcal F)] + \Gamma^\Im_{0,\Delta} \Re e \cdot [\mathcal C ^I(\mathcal F) + \Delta\mathcal C ^I(\mathcal F)] \\
&&+ \Gamma^\Im_1 \Re e \cdot [\mathcal C ^I(\mathcal F)] \cdot \cos\phi_{\gamma\gamma} + \Gamma^\Im_2 \Re e \cdot [\mathcal C ^I(\mathcal F^{eff})] \cdot \cos2\phi_{\gamma\gamma}
\end{eqnarray} 

\noindent where the $\Gamma$s are kinematic coefficients, $\mathcal C ^I(\mathcal F)$, $\Delta\mathcal C ^I(\mathcal F)$ and $\mathcal C ^I(\mathcal F^{eff})$
are respectively twist-2, twist-2 and twist-3 Compton Form Factors (CFFs), as defined in \cite{Bel1} and represent convolution integrals of GPDs. Note that the
cross section difference being only sensitive to the imaginary part of CFFs, can be written as a linear combination of GPDs evaluated at $x=\pm\xi$. Fig.~3 shows
the cross section difference (top) and total cross section (bottom) at the highest $Q^2$ setting of experiment E00-110, along with their respective harmonic
decomposition up to twist-3.
\begin{figure}
   \begin{center}  \includegraphics[width=.8\textwidth]{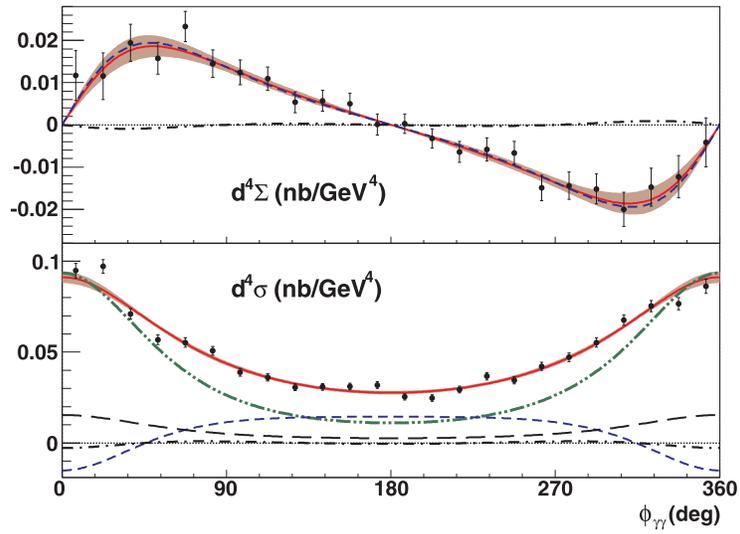}\end{center}
  \caption{E00-113 data (black points) and fits (various curves and error bands): The top plot represents the cross section difference as a function of
           $\phi_{\gamma\gamma}$. A fit to the twist-3 contribution is shown as the short-dashed curve and is very small compared to the main twist-2 contribution in black.
           The bottom plot shows the total cross section as a function of $\phi_{\gamma\gamma}$. The Bethe-Heitler contribution is shown in green. The twist-2 interference
           contributions are the blue and long-dashed curves. Again, the twist-3 contribution is the short-dashed curve and is very strongly suppressed.}
\end{figure}

The main result of E00-110 is two-fold: firstly, the twist-3 contribution to both the difference of cross sections and the total cross section were found small
compared to the twist-2 contributions, which is expected if indeed the twist-2 handbag diagram is dominant. Secondly, the $\Im m \cdot [\mathcal C ^I(\mathcal F)]$ coefficient
extracted from the cross section difference was found to be independent of $Q^2$, which again is the sign that no higher order corrections enter this extracted coefficient.
The conclusion from this study is that E00-110 found the handbag twist-2 contribution dominating (i.e. scaling) even at $Q^2$ of the order 2~GeV$^2$, much like the situation
in regular Deep Inelastic Scattering where scaling is observed at low $Q^2$ as well.

Another interesting result comes from the measurement of the total cross section: it seems very unlikely that the interference term is responsible for the difference
between the Bethe-Heitler cross section and the full electroproduction cross section. Therefore, the DVCS contribution to the cross section might be significant even
in the Jefferson Lab energy range.

\section{Hall B experiment E01-113}

The first dedicated DVCS experiment ran in Hall B in the spring of 2005, using an upgraded CLAS spectrometer \cite{NimB}.
Compared to previous results, the main difference is that a complete three-particle
final state was required for the event to be used for analysis, ensuring a much better exclusivity and less contamination issues. In order to increase the photon
acceptance, a new inner calorimeter was added, centered around the beamline,
at 55~cm from the target and covering the angles between 5 and 15$^\circ$. This calorimeter was built from 424 lead tungstate crystals of quasi-pyramidal shape,
read by avalanche photodiodes (APDs). Since this calorimeter is at such low angles, it was necessary to shield it from Moeller electrons by using a custom-designed
two-coil supraconducting solenoid surrounding the liquid hydrogen target and focusing the Moeller electrons in the central hole of the new calorimeter.

After careful selection of the $(e,p,\gamma)$ final state, the $e p \to e p \gamma$ events show up very clearly in a missing energy spectrum as shown on Fig.~4.
The residual $\pi^0$ contribution was removed using a technique similar to the Hall A experiment: two-photon-decay $\pi^0$ were selected in the data,
and the one-photon-detected asymmetric decay was infered using the ratio of acceptances given by a Monte-Carlo. After $\pi^0$ subtraction, the events were binned
in $x_B$, $Q^2$, $t$ and $\phi_{\gamma\gamma}$. Preliminary results on the asymmetry as a function of $x_B$, $Q^2$, $t$ and $\phi_{\gamma\gamma}$ were shown at the
conference but are not presently available for circulation \cite{FX}. The data were compared where possible with previous measurements and Hall A data points, and were found
compatible. A parametrization by VGG shows a  reasonable agreement \cite{VGG1,VGG2}, especially at high $t$.
The low-$t$ behavior is not as easy to reproduce, both in shape and amplitude. Even though
the addition of a D-term in the VGG parametrization has a rather large and poorly understood impact on the beam spin asymmetry, it is found to
significantly improve the agreement with our measurement.
\begin{figure}
   \begin{center}  \includegraphics[width=.5\textwidth]{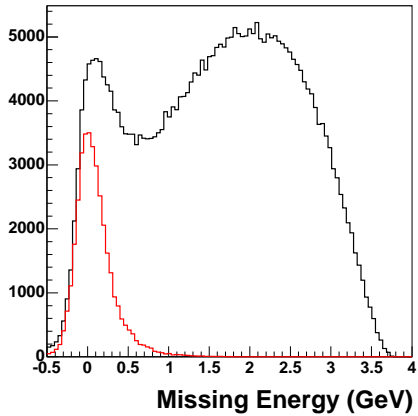}\end{center}
  \caption{Histogram of the missing energy in the reaction $e p \to e p \gamma X$ in the E01-113 experiment, using only the inner calorimeter for the
           photon detection. In black, all configurations are shown, whereas in red, only the configurations passing kinematical cuts on the transverse momenta
           and photon angles for the $e p \to e p \gamma $ are shown. The exclusive peak appears clearly and therefore allows for an unambiguous selection of
           exclusive events.}
\end{figure}

This data set represents a huge improvement both in statistics and kinematical coverage with respect to
previous studies of the beam spin asmmetry. It will clearly have an important impact on GPD models and parametrizations.

\section{Summary and outlook}

Since the mid 90's, both the theoretical and experimental activities related to GPDs have been intense. It is clear now that extraction of GPDs from data
is possible, and that it leads to a rich phenomenology yielding brand new information on the nucleon structure. Deeply Virtual Compton Scattering is clearly
the golden process to start this systematic study of GPDs: even though it is experimentally not the easiest, especially because of the $\pi^0$ contamination
which needs to be dealt with, it is theoretically much cleaner than meson production, and offers a direct access to the imaginary part of the Compton Form Factors,
and therefore GPDs evaluated at $x=\pm\xi$.

The Hall A E00-110 data clearly demonstrates that scaling is already in progress at $Q^2$ around 2~GeV$^2$. This is not such a surprise considering the similarities
of DVCS with regular Deep Inelastic Scattering, which shows similar scaling properties at the same $Q^2$ value. In addition, the Hall A experiment extracted the first 
model-independent linear combination of GPDs. The Hall B E01-113 data measured the beam spin asymmetry over a wide kinematical range in the quark valence region,
and will put strong contraints on GPD models.

Two Hall B experiments at Jefferson Lab will take data in 2008, aiming at collecting even more statistics for the beam spin asymmetry, and measuring the target spin
asymmetry in a wide kinematical range, in order to better contrain the GPD $\widetilde H$. In the longer term, GPD studies is one of the main focus of the 12 GeV
program at Jefferson Lab, which should start around 2013. The first round of GPD-related proposals have been accepted in August 2006, including two DVCS experiments
in Halls A and B, as well as a $\pi^0$ electroproduction experiment. These future sets of data promise to yield results of exceptional accuracy and quality over an
even wider kinematical range.

\section{Acknowledgements}

I acknowledge the help of M.~Gar\c{c}on and F.-X.~Girod in gathering information about the E01-113 experiment in Hall B.

\bibliographystyle{aipproc}   

\bibliography{spin06_sabatie}

\IfFileExists{\jobname.bbl}{}
 {\typeout{}
  \typeout{******************************************}
  \typeout{** Please run "bibtex \jobname" to optain}
  \typeout{** the bibliography and then re-run LaTeX}
  \typeout{** twice to fix the references!}
  \typeout{******************************************}
  \typeout{}
 }

\end{document}